\documentclass[twocolumn,letter]{jpsj3}

\title{Nonlinear Conduction by Melting of Stripe-Type Charge Order in
Organic Conductors with Triangular Lattices
}

\author{
\name{Yasuhiro \surname{Tanaka}}\thanks{E-mail address: yasuhiro@ims.ac.jp}
and \name{Kenji \surname{Yonemitsu}}}

\inst{
Institute for Molecular Science, Okazaki, Aichi 444-8585, Japan\\
Department of Functional Molecular Science, Graduate University for
Advanced Studies, Okazaki, Aichi\ 444-8585, Japan\\
JST, CREST, Tokyo 102-0075, Japan}
      
\abst{
We theoretically discuss the mechanism for the peculiar nonlinear
 conduction in
 quasi-two-dimensional organic conductors $\theta$-(BEDT-TTF)$_2$X 
[BEDT-TTF=bis(ethylenedithio)tetrathiafulvalene] through the melting of
 stripe-type charge order. 
An extended Peierls-Hubbard model attached to metallic electrodes is
 investigated by a nonequilibrium Green's function technique. A novel
 current-voltage characteristic appears in a coexistent state of
 stripe-type and nonstripe 3-fold charge orders, where the applied bias
 melts mainly the stripe-type charge order through the reduction of lattice
 distortion, whereas the 3-fold charge order survives. These contrastive
 responses of the two different charge orders are consistent with the
 experimental observations.
}

\kword{nonlinear conduction, charge order, nonequilibrium Green's
function, organic conductor}

\begin{document}
\maketitle

Nonlinear conduction in low-dimensional electron systems has been of
great interest from the viewpoint of fundamental nonequilibrium physics
and possible applications to electronic devices.
A well-known example is a sliding of density waves in
quasi-one-dimensional materials, where a nesting
of the Fermi surface is responsible for their ground
states\cite{Gruner_94}. In strongly correlated systems such as Mott
insulators\cite{Taguchi_PRB00} and charge-ordered states of
transition metal oxides\cite{Yamanouchi_PRL99}, 
dielectric breakdown phenomena have been observed. In one dimension, a
breakdown of Mott insulators by the Landau-Zener tunneling mechanism has
been proposed theoretically\cite{Oka_PRL03} and the relevance to
experimental findings has been discussed so far.

\begin{figure}
\begin{center}
\includegraphics[height=2.7cm]{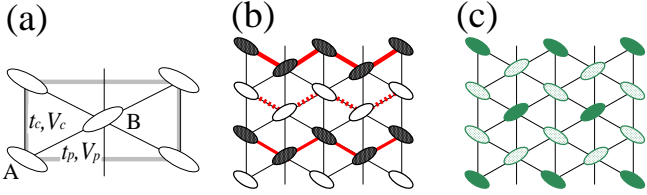}
\end{center}
\caption{\label{fig:fig1}(Color online). (a) Schematic view of the
 high-temperature structure for $\theta$-(ET)$_2$X. Ellipses represent
 ET molecules. The gray rectangle shows a unit cell in
 which sites are labeled $A$ and $B$. (b)
 Horizontal CO and (c) 3-fold CO, where the solid
 (open or shaded) ellipses denote hole-rich (-poor) molecules. In (b),
 the lattice distortion along the horizontal stripe is
 shown by the thick and broken lines. }
\end{figure}

The observations of giant nonlinear conduction and spontaneous current
oscillation in the organic compounds
$\theta$-(BEDT-TTF)$_2$Cs$M$(SCN)$_4$\cite{Sawano_NATURE05,Suko_Mat10}
($M$=Co and Zn) have renewed interest since their mechanism and the
electric-field-induced behaviors seem to differ in many respects from
those in the above materials. 
The family of organic conductors $\theta$-(ET)$_2$X (ET is the
abbreviation of BEDT-TTF) is known to
exhibit charge order (CO)\cite{Miyagawa_PRB00,Chiba_JPCS01}. It has a
quasi-two-dimensional structure, where ET molecules form a triangular
lattice in each conduction layer
[Fig. \ref{fig:fig1}(a)]\cite{Mori_PRB98} with electron density at 3/4
filling (one hole per two ET molecules). 
In $\theta$-(ET)$_2$RbZn(SCN)$_4$, a metal-insulator transition
with a structural distortion occurs at $T_{\rm
CO}=190$ K\cite{Mori_PRB98}. A
stripe-type arrangement of localized charges indexed by wave number
${\bf q}_2=(0,0,1/2)$, which is called a horizontal CO
[Fig. \ref{fig:fig1}(b)], emerges below $T_{\rm
CO}$\cite{Tajima_PRB00,Yamamoto_PRB02,M_Watanabe_JPSJ04}. In
$\theta$-(ET)$_2$Cs$M$(SCN)$_4$, on the other hand, coexistence of
two different COs, the horizontal CO and a nonstripe CO indexed by 
${\bf q}_1=(2/3,k,1/3)$, has been observed by X-ray
experiments\cite{Watanabe_JPSJ99,Nogami_SynMet99}, although this
compound does not show any long-range order.
Several experiments suggest that the peculiar nonlinear conduction
results from suppression of the horizontal CO by an electric field
without destroying the nonstripe CO\cite{Sawano_NATURE05,Ito_EPL08}. A
similar nonlinear conduction and coexistence of two kinds of COs
have also been observed in a rapidly cooled RbZn
salt\cite{Inada_PRB09,Nogami_JPSJ10}. 
Although nonlinear conduction has been found in other
compounds\cite{Mori_PRL08,Mori_PRB09}, this unique feature of multiple
charge modulation seems essential for the nonlinearity in
$\theta$-type compounds.

Theoretically, the CO phenomenon in $\theta$-(ET)$_2$X has been
investigated from various aspects\cite{Seo_JPSJ06,Kuroki_SciTec09}. The
charge disproportionation results mainly from the long-range nature
of the Coulomb interaction, whereas electron-phonon (e-ph) couplings are
also important. In particular, the
lattice distortion in the RbZn salt considerably stabilizes the horizontal
CO\cite{Tanaka_JPSJ07,Miyashita_PRB07,Tanaka_JPSJ08}. Compared with our
knowledge on the ground states, nonequilibrium states induced by an
external field have been poorly
understood. Recently, Mori and coworkers have shown that a
phenomenological equation can reproduce the
observed nonlinear current-voltage characteristics in some
compounds\cite{Mori_PRL08,Mori_PRB09}. However, the origin of nonlinear
conduction is still unclear, so a microscopic theory is highly
desirable.

With these in mind, we investigate nonequilibrium steady states of
$\theta$-(ET)$_2$X under applied bias voltages, using a model that takes
account of both the long-range Coulomb interactions and e-ph
couplings\cite{Tanaka_JPSJ07,Miyashita_PRB07,Tanaka_JPSJ08}. The
model describes competition among various COs, including the
horizontal CO and a so-called 3-fold CO [Fig. \ref{fig:fig1}(c)], the
latter has a nonstripe charge pattern\cite{Mori_JPSJ03}
and can be related to the CO with ${\bf q}_1$ in the Cs$M$ salt. 
The state with the horizontal CO is insulating, whereas that with the
3-fold CO is metallic\cite{Kaneko_JPSJ06}. We show that when these two
states coexist, the bias voltage melts the horizontal CO and largely
alters the conduction behavior. The lattice distortion, which induces
the horizontal charge modulation, has a key role in determining whether
the system becomes resistive or conductive. 

We consider the extended Peierls-Hubbard
model\cite{Tanaka_JPSJ07,Miyashita_PRB07,Tanaka_JPSJ08} written as
\begin{eqnarray}
H&=&\sum_{\langle ij \rangle \sigma} \left[
(t_{i,j} + \alpha_{i,j} u_{i,j}) c^\dagger_{i\sigma} c_{j\sigma} +\mbox{h.c.}
\right] \nonumber \\ 
&+&U\sum_i (n_{i\uparrow}-N_e/2)(n_{i\downarrow}-N_e/2)\nonumber \\
&+&\sum_{\langle\langle ij \rangle\rangle} V_{i,j}(n_i-N_e)(n_j-N_e)
+\sum_{\langle ij \rangle } \frac{K_{i,j}}{2} u_{i,j}^2\ ,
\label{eq:ham}
\end{eqnarray}
where $\langle ij\rangle$ represents the summation over pairs of
neighboring sites, $c^{\dagger}_{i\sigma}(c_{i\sigma})$ denotes the
creation (annihilation) operator for an electron with spin $\sigma$ at
the $i$th site, $n_{i\sigma}=c^{\dagger}_{i\sigma}c_{i\sigma}$, 
$n_{i}=n_{i\uparrow}+n_{i\downarrow}$, and the averaged electron density
$N_e=1.5$. 
$t_{i,j}$ denotes the transfer integrals and $U$ the on-site repulsion. 
For the intersite Coulomb interactions $V_{i,j}$, we consider up to
third-neighbor pairs of sites, the summation over which is represented by
$\langle \langle ij\rangle\rangle$. The e-ph coupling constant, the
lattice displacement and the elastic constant
are denoted by $\alpha_{i,j}$, $u_{i,j}$, and $K_{i,j}$,
respectively. We introduce new variables as
$y_{i,j}=\alpha_{i,j}u_{i,j}$ and
$s_{i,j}=\alpha_{i,j}^{2}/K_{i,j}$\cite{Tanaka_JPSJ07,Miyashita_PRB07,Tanaka_JPSJ08}. 

The structure of $\theta$-(ET)$_2$X in the high-temperature metallic
phase is shown in Fig. \ref{fig:fig1}(a). There are two transfer
integrals $t_c$ and $t_p$ on the vertical and diagonal bonds,
respectively. We set $t_c=-0.04$ (eV) and $t_p=0.1$ in the following. On
the vertical (diagonal) bonds, we define the nearest-neighbor
interaction $V_c$ ($V_p$). For the second- and
third-neighbor interactions, we write them as
$V_{i,j}=V_{lr}/r_{ij}$. Here, $r_{ij}$ is the distance between the
$i$th and $j$th sites. The horizontal CO and the 3-fold CO are
schematically shown in Figs. \ref{fig:fig1}(b) and 1(c), respectively. 
For e-ph couplings, we consider a lattice distortion caused by the
molecular rotation, which is crucial for stabilizing the horizontal
CO\cite{Tanaka_JPSJ07,Miyashita_PRB07,Tanaka_JPSJ08}. This is because it
gives homogeneous modulation in the transfer integrals 
along with a horizontal stripe. We denote the modulation as
$y_{\phi}$ and
write the corresponding e-ph coupling as $s_{\phi}$. These variables are
defined only on the bonds that are depicted by the thick and broken lines in
Fig. \ref{fig:fig1}(b). We assume that $y_{\phi}$ is independent of bond
index so that the transfer integrals on the hole-rich (-poor) stripe
are written as $t_p+y_{\phi}$ ($t_p-y_{\phi}$) with $y_{\phi}>0$. The
elastic energy for each distorted bond is given by
$y_{\phi}^2/(2s_{\phi}$). For simplicity, we do not take account of
other types of
modulation\cite{Tanaka_JPSJ07,Miyashita_PRB07,Tanaka_JPSJ08}. 

\begin{figure}
\begin{center}
\includegraphics[height=3.5cm]{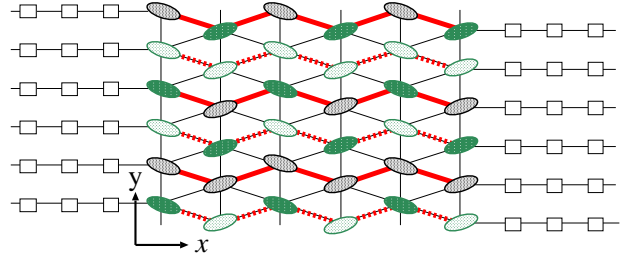}
\end{center}
\caption{\label{fig:fig2}(Color online). Schematic picture of the
 model. The left and right electrodes are attached to the central part
 where the horizontal and 3-fold COs coexist when the bias voltage is
 absent. A horizontal charge modulation induced
 by the lattice distortion on the 3-fold charge pattern is represented
 by the gray ellipses on the thick bonds. The $x$-axis ($y$-axis) is
 along (perpendicular to) the conduction direction.}
\end{figure}

We describe the steady states under applied bias voltages by the
nonequilibrium Green's function
method\cite{Yonemitsu_JPSJ09,Tanaka_PRB11}.  As shown in
Fig. \ref{fig:fig2}, we attach the left and right
$(\alpha=L,R)$ semi-infinite metallic electrodes to the
central part that is described by eq. (\ref{eq:ham}). A coexistent CO,
which will be
discussed later, is realized in the absence of the bias voltage.
For the $i$th site [$i=(i_x,i_y)$] in the central part, we define its
coordinates as ($i_x$, $i_y$) if $i_x$ is odd, and ($i_x$, $i_y-1/2$) if
$i_x$ is even. The numbers of sites are denoted by $L_x$ and $L_y$
($1\leq i_x\leq L_x$ and $1\leq i_y\leq L_y$). 
We assume that electrons in the leads are noninteracting and that 
they move only in the $x$-direction; for simplicity, the electrodes are
one-dimensional. The effects of the leads on the central part are
incorporated into
the self-energies. In the wide-band limit, the retarded self-energies
are independent of energy and written as 
$(\Sigma^r_{\alpha})_{ij}=-\frac{i}{2}\gamma_{\alpha}\sum_{i_{\alpha}}
\delta_{ii_{\alpha}}\delta_{ji_{\alpha}}$, where $\delta_{ij}$ is the
Kronecker delta, $i_L$ ($i_R$) denotes the site that is connected
with the left (right) electrode, and $\gamma_{\alpha}$ the coupling
constant between the central part and the electrode $\alpha$.

We use the Hartree-Fock approximation for the $U$, $V_p$, and $V_c$ terms in
eq. (\ref{eq:ham}). For the second- and third-neighbor interactions, we
employ the Hartree approximation.
These interactions contribute to a redistribution of charges near the
electrodes\cite{Tanaka_PRB11}, whereas the charge disproportionation is
mainly caused by $V_p$ and $V_c$. 
We consider the case of $V_p,\ V_c\gg V_{lr}$. 
The periodic boundary condition is adopted along the
$y$-axis. In the mean-field Hamiltonian, we introduce a scalar potential
$\psi$ that is defined by the Hartree terms for the intersite Coulomb
interactions as 
$\psi(i_x,i_y)=\sum_{j\neq i}V_{i,j}(\langle n_j\rangle-N_e)+ai_x+b$,
which is equivalent to the Poisson equation\cite{Yonemitsu_JPSJ09}. 
The slope of the potential, $ai_x$, describes the electric field in the
central part.\cite{Tanaka_PRB11}
The constants $a$ and $b$ are so determined that $\psi$ satisfies the
boundary conditions, $\frac{1}{L_y}\sum_{i_y}\psi(1,i_y)=V/2$
and $\frac{1}{L_y}\sum_{i_y}\psi(L_x,i_y)=-V/2$, when the bias voltage
$V$ is applied to the system. We assume that the work-function
differences at the interfaces are absent. 

The steady states under the applied bias are obtained as
a self-consistent solution for the mean fields that are calculated by
the method described in ref. 27.
 The electron density $\langle n_{i\sigma}\rangle$ is calculated by
decomposing it into the ``equilibrium'' and ``nonequilibrium'' parts
as $\langle n_{i\sigma}\rangle =n^{\rm eq}_i+\sum_{\alpha}\delta
n^{\alpha}_{i\sigma}$\cite{Yonemitsu_JPSJ09}. The same decomposition is
used to obtain $\langle c^{\dagger}_{i\sigma}c_{j\sigma}\rangle$. 
We adjust the chemical
potential $\mu_C=(\mu_L+\mu_R)/2$\cite{Yonemitsu_JPSJ09} such that
the electron density of the central part is fixed at 3/4 filling. Here,
$\mu_L$ and $\mu_R$ are the left and right chemical potentials,
respectively. For finite $V$, we set $\mu_L=\mu_C+V/2$ and
$\mu_R=\mu_C-V/2$. The lattice distortion $y_{\phi}$ is determined as in
the equilibrium case\cite{Tanaka_JPSJ07}. The current $J$ is 
obtained by using $\delta n^{\alpha}_{i\sigma}$ as
$J=\gamma_R\sum_{i_R\sigma}\delta
n^{L}_{i_R\sigma}-\gamma_L\sum_{i_L\sigma}\delta
n^{R}_{i_L\sigma}$\cite{Yonemitsu_JPSJ09}, where we set $e=\hbar=1$. 
In the following, we use $U=0.6$, $V_c/U=0.35$, $s_{\phi}=0.1$, and
$V_{lr}=0.02$. The size of the central part is $L_x=L_y=18$ unless
otherwise noted.

\begin{figure}
\begin{center}
\includegraphics[height=6.5cm]{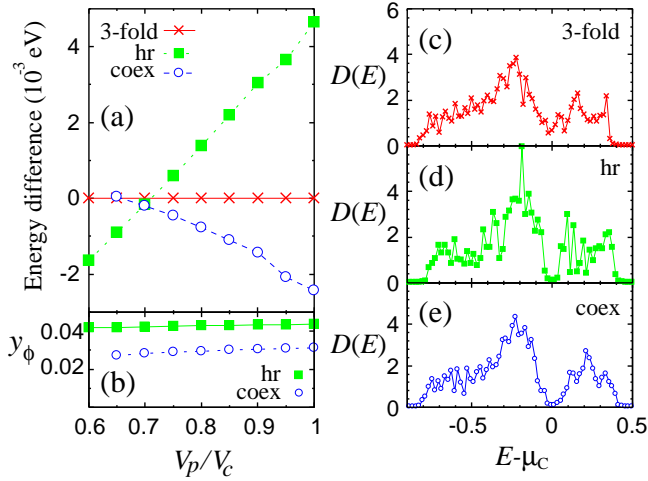}
\end{center}
\caption{\label{fig:fig3}(Color online). (a) Energies of the horizontal
 (hr) and coexistent (coex) COs with $L_x=L_y=18$, $U=0.6$,
 $V_c/U=0.35$, $s_{\phi}=0.1$, $V_{lr}=0.02$, and $\gamma_R=\gamma_L=0$,
 relative to that of the 3-fold CO, as a function of $V_p/V_c$. (b) The
 lattice distortions $y_{\phi}$
 for the horizontal and coexistent COs. Density of
 states for (c) the 3-fold, (d) horizontal, and (e) coexistent COs in
 the case of $V_p/V_c=1$.}
\end{figure}

First, we consider the equilibrium case where the CO system is isolated
from the electrodes ($\gamma_R=\gamma_L=0$). 
The ground-state energies per site of three mean-field solutions as a
function of $V_p/V_c$ are shown in Fig. \ref{fig:fig3}(a), where the
energy of the 3-fold CO is set at zero. The horizontal and coexistent
COs have a finite lattice distortion $y_{\phi}$, as shown in
Fig. \ref{fig:fig3}(b). For $V_p/V_c<0.7$, the horizontal CO is the most
stable, whereas the 3-fold CO has a lower energy than the horizontal CO
for $V_p/V_c>0.7$. This results from charge frustration on the
triangular lattice\cite{Mori_JPSJ03,Kaneko_JPSJ06}. 
The 3-fold CO is further stabilized by coexisting
with the $y_{\phi}$-induced horizontal CO (Fig. \ref{fig:fig2}) and
becomes the ground state near $V_p/V_c=1$\cite{Tanaka_JPSJ07}. 
For each CO pattern, we show the density of states
$D(E)$ in Figs. \ref{fig:fig3}(c)-\ref{fig:fig3}(e) for $V_p/V_c=1$,
where we used a broadening factor of $\eta=0.01$. A finite
$D(E)$ at the Fermi level exists for the 3-fold CO since this state is
metallic\cite{Kaneko_JPSJ06}, whereas the horizontal CO has an energy
gap. In the coexistent CO, $D(E)$ at $E=\mu_C$ is suppressed compared
with that for the 3-fold CO, which is due to the horizontal CO. 
However, we note that the state has no energy gap in the thermodynamic
limit\cite{Tanaka_JPSJ07}.

\begin{figure}
\begin{center}
\includegraphics[height=6.5cm]{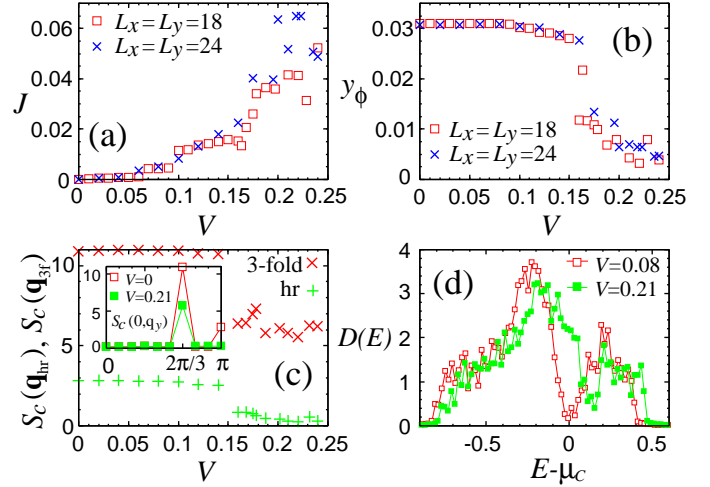}
\end{center}
\caption{\label{fig:fig4}(Color online). (a) Electric current $J$, (b)
 lattice distortion $y_{\phi}$, and (c) charge structure factors 
$S_c({\bf q}_{\rm hr})$ and $S_c({\bf q}_{\rm 3f})$ as a function of
 $V$, where ${\bf q}_{\rm hr}$ and ${\bf q}_{\rm 3f}$ are the wave
 vectors for the horizontal (hr) and 3-fold COs, respectively. In (a)
 and (b), results with $L_x=L_y=24$ are also shown. In (c), the $q_y$
 dependence of $S_c(0,q_y)$ is shown in the inset. 
(d) Density of states for $V=0.08$ and $0.21$.}
\end{figure}

Next, we discuss the results obtained with finite bias $V$. We set
$V_p/V_c=1$ and $\gamma_L=\gamma_R=0.03$, where the ground
state without the bias is the coexistent CO. The current-voltage
characteristics and the distortion
$y_{\phi}$ are shown in Figs. \ref{fig:fig4}(a) and \ref{fig:fig4}(b),
respectively. In these figures, we show the results with
$L_x=L_y=24$ for comparison. The behaviors of $J$ and $y_{\phi}$ are
qualitatively the same as those with $L_x=L_y=18$. With increasing $V$,
the current gradually increases and abruptly becomes
large at $V_{cr}\sim 0.17$. For small $V$ and $L_x=L_y=18$, $J$ has
stepwise structures
owing to the finite-size effect\cite{Tanaka_PRB11}. As shown in
Fig. \ref{fig:fig4}(b), $y_{\phi}$ is
almost unchanged for $V<V_{cr}$, although it shows a gradual
decrease with increasing $V$. At $V=V_{cr}$, $y_{\phi}$ steeply decreases,
which is directly related to the reduction of the horizontal CO. 

The effects of bias voltages on the horizontal and
3-fold components of the charge distribution are obtained by calculating
the charge structure factor, which is defined as
\begin{equation}
S_c({\bf q})=\frac{1}{N_s}\sum_{\mu,\nu}(\langle n_{\mu A}n_{\nu
A}\rangle+\langle n_{\mu B}n_{\nu B}\rangle)e^{i{\bf q}({\bf
R}_{\mu}-{\bf R}_{\nu})}.
\end{equation}
Here, we use the unit cell shown in Fig. \ref{fig:fig1}(a), which is
labeled by $\mu$, $\nu$. $A$ and $B$ are indices for sites inside the
unit cell. The position vector for the $\mu$-th ($\nu$-th) unit cell
is denoted by ${\bf R}_{\mu}$ (${\bf R}_{\nu}$) and $N_s=L_x\times L_y$.
The wave vectors that correspond to the horizontal and 3-fold
components are ${\bf q}_{\rm hr}=(0,\pi)$ and ${\bf
q}_{\rm 3f}=(0,2\pi/3)$, respectively. In Fig. \ref{fig:fig4}(c), we
show $S_c({\bf q}_{\rm hr})$ and $S_c({\bf q}_{\rm 3f})$ as a function
of $V$, and the $q_y$ dependence of $S_c(0,q_y)$ for $V=0$ and $0.21$ in
the inset. In the $V=0$ case, $S_c(0,q_y)$ has two peaks at ${\bf q}_{\rm
hr}$ and ${\bf q}_{\rm 3f}$ since the two COs coexist. Because
the 3-fold charge modulation is larger than the horizontal one, we have 
$S_c({\bf q}_{\rm 3f})>S_c({\bf q}_{\rm hr})$. For
$V<V_{cr}$, the values of $S_c({\bf q}_{\rm hr})$ and $S_c({\bf q}_{\rm
3f})$ are almost unchanged, which indicates that the coexistent CO is
robust against the applied bias.
For $V>V_{cr}$, both $S_c({\bf q}_{\rm hr})$ and
$S_c({\bf q}_{\rm 3f})$ decrease and the charge distribution is largely
modified. In particular, the horizontal component is drastically
weakened. For $V=0.21$, the ${\bf q}_{\rm hr}$ peak in $S_c(0,q_y)$
disappears. Correspondingly, $y_{\phi}$ becomes very small for large
$V$. However, the 3-fold component survives even in the region
$V>V_{cr}$. It is noteworthy that $J$ increases without destroying the
3-fold CO. We show the density of states for
$V=0.08$ and $0.21$ in Fig. \ref{fig:fig4}(d). For $V=0.08$,
$D(E)$ is qualitatively the same as that in the $V=0$ case shown in
Fig. 3(e). Since the coexistent CO has no energy
gap\cite{Tanaka_JPSJ07}, a small current can flow even for $V<V_{cr}$.
For $V=0.21$, a large $D(E)$ appears at $E=\mu_C$ since the horizontal
charge modulation is suppressed. The change in the conduction behavior
is triggered by the reduction in $y_{\phi}$. If $y_{\phi}$ is large, the
horizontal CO persists so the system is resistive, whereas if $y_{\phi}$
decreases, only the 3-fold CO remains so the system becomes conductive. 

Let us discuss the relevance to experimental results on
$\theta$-(ET)$_2$X. Our results basically reproduce the X-ray
results\cite{Sawano_NATURE05,Nogami_JPSJ10,Ito_EPL08} that indicate that
the nonlinear conduction is caused by the melting of the horizontal CO
whereas the nonstripe CO remains. In the CsZn salt, the resistivity
begins to increase at around 50 K\cite{Mori_PRB98}. This corresponds to
the growth of X-ray
intensity for the horizontal CO, whereas that for the nonstripe CO
shows only slight temperature
dependence\cite{Watanabe_JPSJ99,Nogami_SynMet99}. Theoretically,
such contrastive temperature dependences of stripe-type and 3-fold-type
charge fluctuations have been shown by the random
phase approximation\cite{Udagawa_PRL07}, where the former
comes from the Fermi-surface instability assisted by e-ph couplings,
whereas the latter is due to the wave-vector dependence of the Fourier
transform of the intersite Coulomb interaction. These facts also
suggest that the
horizontal CO induced by the lattice distortion is directly related to
the resistive behavior, which is consistent with our results. However,
there are some issues that remain to be clarified. Experimentally, the
nonlinear conduction appears in a state with no long-range CO, which is
in contrast to the results of our mean-field calculations. Therefore, at
present, it is difficult to compare the
results quantitatively. In fact, the electric fields required for
nonlinearity ($\sim$1 V/cm for the CsZn salt
and $\sim$10 V/cm for the rapidly cooled RbZn salt) are much smaller than
that in the present study. For a quantitative comparison, the effects of
quantum fluctuations must be taken into account. The finite-size
effects\cite{Tanaka_PRB11} as well as the thermal fluctuations will
affect the values of the threshold voltage.
Recently, an inhomogeneous state of competing COs has
been suggested and a possible relation to the nonlinear conduction has
been discussed
\cite{Sawano_NATURE05,Nogami_JPSJ10,Ito_EPL08,Inada_PRB09}. 
Although the present study is based on a uniform CO, we speculate that 
in a spatially nonuniform state, only domains of the horizontal CO are
suppressed by an electric field, which results in nonlinear
current-voltage characteristics similar to our results. The origin of
the current oscillation in $\theta$-(ET)$_2$X\cite{Sawano_NATURE05}
occurring with the nonlinear conduction 
is still unclear. However, Suko {\it et al.}\cite{Suko_Mat10} have
recently suggested that the oscillation is due to a current-induced
modulation of the lattice distortion, which may be related to our
results. 

In summary, we have investigated the mechanism of 
nonlinear conduction in $\theta$-(ET)$_2$X. In the coexistent state of
horizontal and 3-fold COs, the bias voltage weakens the
lattice distortion and melts the horizontal CO. The metallic 3-fold CO
remains even after the disappearance of the horizontal CO, which leads
to selective melting of the latter CO. We have shown that these
different responses depending on the spatial patterns of the two COs
bring about the novel nonlinear conduction. 

\begin{acknowledgments}
This work was supported by Grants-in-Aid for Scientific Research (C) (Grant
No. 23540426), Scientific Research (B) (Grant No. 20340101) and Scientific
Research (A) (Grant No. 23244062), and by 
``Grand Challenges in Next-Generation Integrated Nanoscience''
from the Ministry of Education, Culture, Sports, Science and Technology
 of Japan.
\end{acknowledgments}

\end{document}